\documentclass{aa}  
\usepackage{graphicx}
\usepackage{txfonts}
\begin{document}

%\input{psfig.sty}
%
% Next 5 lines define \simless and \simgreat: "less than or approximately
% equal to" and "greater than or approximately equal to".
\newbox\grsign \setbox\grsign=\hbox{$>$} \newdimen\grdimen \grdimen=\ht\grsign
\newbox\simlessbox \newbox\simgreatbox
\setbox\simgreatbox=\hbox{\raise.5ex\hbox{$>$}\llap
     {\lower.5ex\hbox{$\sim$}}}\ht1=\grdimen\dp1=0pt
\setbox\simlessbox=\hbox{\raise.5ex\hbox{$<$}\llap
     {\lower.5ex\hbox{$\sim$}}}\ht2=\grdimen\dp2=0pt
\def\simgreat{\mathrel{\copy\simgreatbox}}
\def\simless{\mathrel{\copy\simlessbox}}
% Next lines define "approximately proportional to"
\newbox\simppropto
\setbox\simppropto=\hbox{\raise.5ex\hbox{$\sim$}\llap
     {\lower.5ex\hbox{$\propto$}}}\ht2=\grdimen\dp2=0pt
\def\simpropto{\mathrel{\copy\simppropto}}

\title{Halo intruders in the Galactic bulge revealed by
HST and Gaia: the globular clusters Terzan~10
 and Djorgovski~1 
\thanks{Observations obtained at the
Hubble Space Telescope, GO-14074 (PI: Cohen), GO-9799 (PI:Rich), and the European Southern Observatory, 
proposals 089.D-0194(A), 091.D-0711(A) (PI: Ortolani).
} }

%\subtitle{CS31082-001}

\author{
S. Ortolani\inst{1}
\and
D. Nardiello\inst{1}
\and
A. P\'erez-Villegas\inst{2}
\and
E. Bica\inst{3}
\and
B. Barbuy\inst{2}
}
\offprints{B. Barbuy (b.barbuy@iag.usp.br).}

\institute{
Universit\`a di Padova, Dipartimento di Fisica e Astronomia Galileo Galilei, 
Vicolo dell'Osservatorio 2, I-35122 Padova, Italy;\\
\and
Universidade de S\~ao Paulo, IAG, Rua do Mat\~ao 1226,
Cidade Universit\'aria, S\~ao Paulo 05508-900, Brazil;\\
\and
Universidade Federal do Rio Grande do Sul, Departamento de Astronomia,
CP 15051, Porto Alegre 91501-970, Brazil;\\
}
 
\date{Received; accepted }
%\abstract{}{}{}{}{} 
% 5 {} token are mandatory
\abstract
{The low-latitude globular clusters Terzan 10 and Djorgovski 1 are 
projected in the Galactic bulge, in a Galactic region highly affected by 
extinction. A discrepancy of a factor of $\sim$2 exists in the literature in regards to the distance determination of these clusters.} 
{We revisit the colour-magnitude diagrams (CMDs) of these two globular
clusters with the purpose of disentangling their
 distance determination ambiguity and, for the first time, of determining 
their orbits to identify whether or not they are part of the bulge/bar region.}
% method
{We use {{\it Hubble Space Telescope}} CMDs,
with the filters F606W from ACS and F160W from WFC3 for Terzan 10, and
F606W and F814W from ACS for Djorgosvski~1, and combine them
 with the proper motions from Gaia Date Release 2. For the orbit integrations, 
we employed a steady Galactic model with bar.
 }
%results
{For the first time the blue horizontal branch of these clusters is clearly
 resolved. We obtain
reliable distances of d$_{\odot}$ = 10.3$\pm$1.0 kpc and 9.3$\pm$0.5 kpc for 
Terzan 10, and Djorgovski 1 respectively, indicating that they
are both currently located in the bulge volume. 
 From Gaia DR 2 proper motions, together with our new distance determination 
and recent literature radial velocities, we are able to show that the
 two sample clusters have typical halo orbits that are passing by the 
bulge/bar region, but that they are not part of this component. 
For the first time, halo intruders are identified in the bulge.}
{}
\titlerunning{Blue horizontal branch bulge globular clusters Terzan 10 and Djorgovski~1}
\authorrunning{S. Ortolani et al.}
\maketitle
%
%________________________________________________________________

\section{Introduction}
The formation of the Galactic bulge is a debated topic within the context
of galaxy evolution (e.g. Renzini et al. 2018). 
Recent studies on low-Galactic-latitude fields
 revealed a complex scenario with a combination of 
different stellar populations including metal-poor and metal-rich
 components with different spatial distribution, kinematics, and
 possibly also distinct ages (Babusiaux et al. 2010, 214).
These studies indicate that the moderately metal-poor population component 
corresponds to a spheroidal old population, 
and the metal-rich one is confined to a bar/disc. 
These two components contain respectively 48\% and 52\% of the stars
according to Zoccali et al. (2018).

The globular clusters (GCs) projected in the direction of the Galactic bulge 
can be used as population tracers and 
can help to constrain the bulge formation and evolution models.
This work is part of an effort to study the globular clusters of the inner bulge
as presented, for example, in Barbuy et al. (1998, 2018) and Bica et al. (2016).

Proper motions derived using data from ground-based and space telescopes, in particular from Gaia, 
 together with orbital calculations for  bulge GCs are
 now becoming available 
(e.g. Casetti-Dinescu 
et al. 2013; Moreno et al. 2014; Rossi et al. 2015; P\'erez-Villegas et al. 2018; Vasiliev 2018). 
These studies show that there are some systematic trends
 in the orbital parameters, possibly related also to  metal abundances, 
indicating ``families'' of GCs with distinct origins.
The present study is dedicated to two moderately metal-poor GCs,
 Terzan 10 and Djorgovski 1, with metallicity 
[Fe/H]$\sim-$1.0 (Ortolani et al. 1995, 1997; Harris 1996, edition 2010).
 These clusters are located at low Galactic latitudes 
%0.7$<$$<|b|>$$<$2.5$^{\circ}$,
 and therefore in very reddened and crowded regions.
 
Terzan~10 (ESO521-SC16),
 discovered by Terzan (1971), is located at J2000
 $\alpha$ = 18$^{\rm h}$02$^{\rm m}$57.8$^{\rm s}$, 
$\delta$ = $-$26$^{\rm o}$$04'01\arcsec$, with Galactic coordinates
l = 4\fdg42, b = $-$1\fdg86. We point out that the coordinates and designations
in Ortolani et al. (1997) were mistyped.

Djorgovski~1, discovered by Djorgovski (1987),
 is located at $\alpha$ = 17$^{\rm h}$47$^{\rm m}$28.7$^{\rm s}$, 
$\delta$ = $-$33$^{\rm o}$$03'59\arcsec$, with Galactic coordinates
l = 356\fdg67, b = $-$2\fdg48.

Recent studies of the bulge area have provided a series of new GCs
 and candidates (Minniti et al. 2018; Camargo 2018;
Ryu \& Lee 2018; Bica et al. 2018; Piatti 2018). 
A consistent derivation of parameters (distance,
reddening, age) requires deep high-resolution images  however, and the  samples already
known are still lacking accurate values for these parameters; a good example is provided by 
the two GCs studied here.
A useful approach that has been leading to newly derived distances is
the systematic observation of variables, in particular in the
VISTA Variables
in the Via Lactea (VVV) survey (Saito et al. 2012). Using RR Lyrae
detected with VVV and other sources, Alonso-Garcia et al. (2015)
derived a distance for Terzan 10. %A combination of accurate CMDs and
%RR Lyrae are being used to derive ages and distances (Kerber et al. 2018a,b).

%The sample cluster parameters are revisited in the present paper using 
%Hubble Space Telescope images, and Gaia data on proper motions.
The main goal of this work is to obtain accurate distances, which, together with
  kinematical data including
recent literature providing radial velocities, and the
 proper motions obtained from Gaia Data Release 2 (DR2)
 (Gaia Collaboration et al. 2018a, 
hereafter Gaia DR2), allow us to reconstruct their orbital parameters.

 Observations are described in Sect. 2, we present Hubble Space Telescope (HST) colour-magnitude diagrams (CMDs) 
with superimposed red giants measured with Gaia in Sect. 3,
cluster parameters and orbits are derived for Terzan 10 and
Djorgovski 1 (hereafter Djorg 1) in Sect. 4, and  conclusions are drawn in   Sect. 5. 
%____________________________ OBSERVATIONS ___________________________
%--- 3 ---

\section{Observations and data reduction}

For this work we used the available {\it HST} observations of the
central regions of the clusters. For Terzan\,10 we made use of the
observations collected during the mission GO-14074 (PI:\,Cohen, mean
epoch 2016.4) in F606W with the Wide Field Channel (WFC) of the
Advanced Camera for Survey (ACS), and in  F160W with the IR
channel of the Wide Field Camera 3 (WFC3). Djorg\,1 data were
collected with ACS/WFC in F606W and F814W filters during the
mission GO-9799 (PI:\,Rich, mean epoch 2004.15).

For the data reduction we adopted the software \texttt{kitchen\_sync2}
described in detail in Nardiello et al. (2018).
% \citet{2018arXiv180904300N}. 
Briefly, using
\texttt{flc} images and perturbed empirical PSF arrays, the software
analyses all the exposures simultaneously to find and measure the
sources. We refer the reader to %\citet{2018arXiv180904300N} 
Nardiello et al. (2018) and Bellini et al. (2017) %\citet{2017ApJ...842....6B}
 for a detailed description of this approach.

Magnitudes have been calibrated into the Vega-mag system by comparing
aperture photometry on \texttt{drc} images against our PSF-fitting
photometry and adopting the photometric zero-points given by the “ACS
Zero-points calculator”\footnote{https://acszeropoints.stsci.edu/} in
the case of ACS/WFC observations, and by Kalirai et al. (2009)
%\citet{2009wfc..rept...30K}
in the case of WFC3/IR data.

\begin{figure}
\includegraphics[angle=-90,width=9cm]{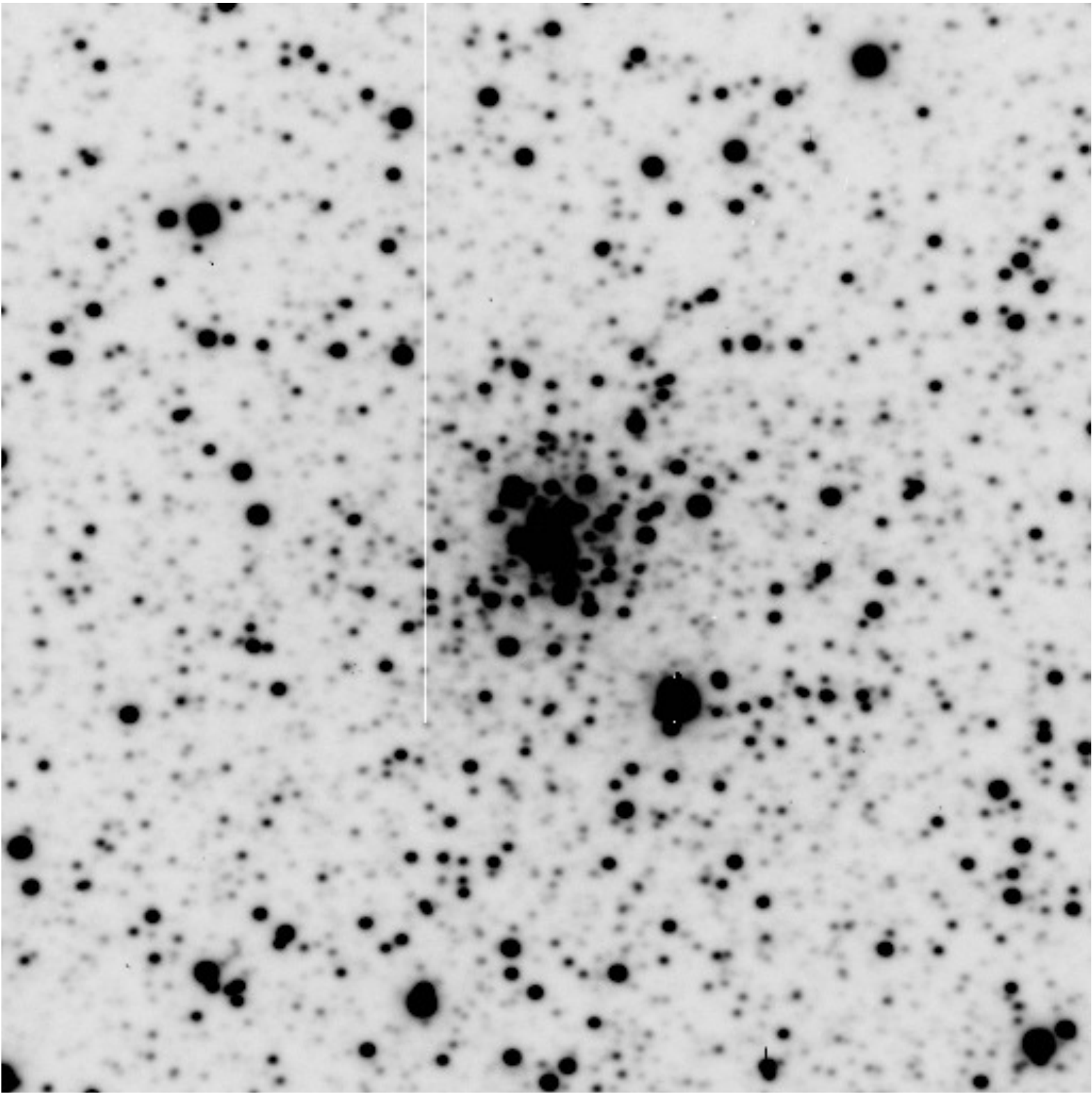}
\caption {NTT 180s I image of Terzan 10. North is up, east to the right.
Size is 2$\times$2 arcmin$^{2}$. }
\label{ter10}
\end{figure}

\begin{figure}
\includegraphics[angle=-90,width=9cm]{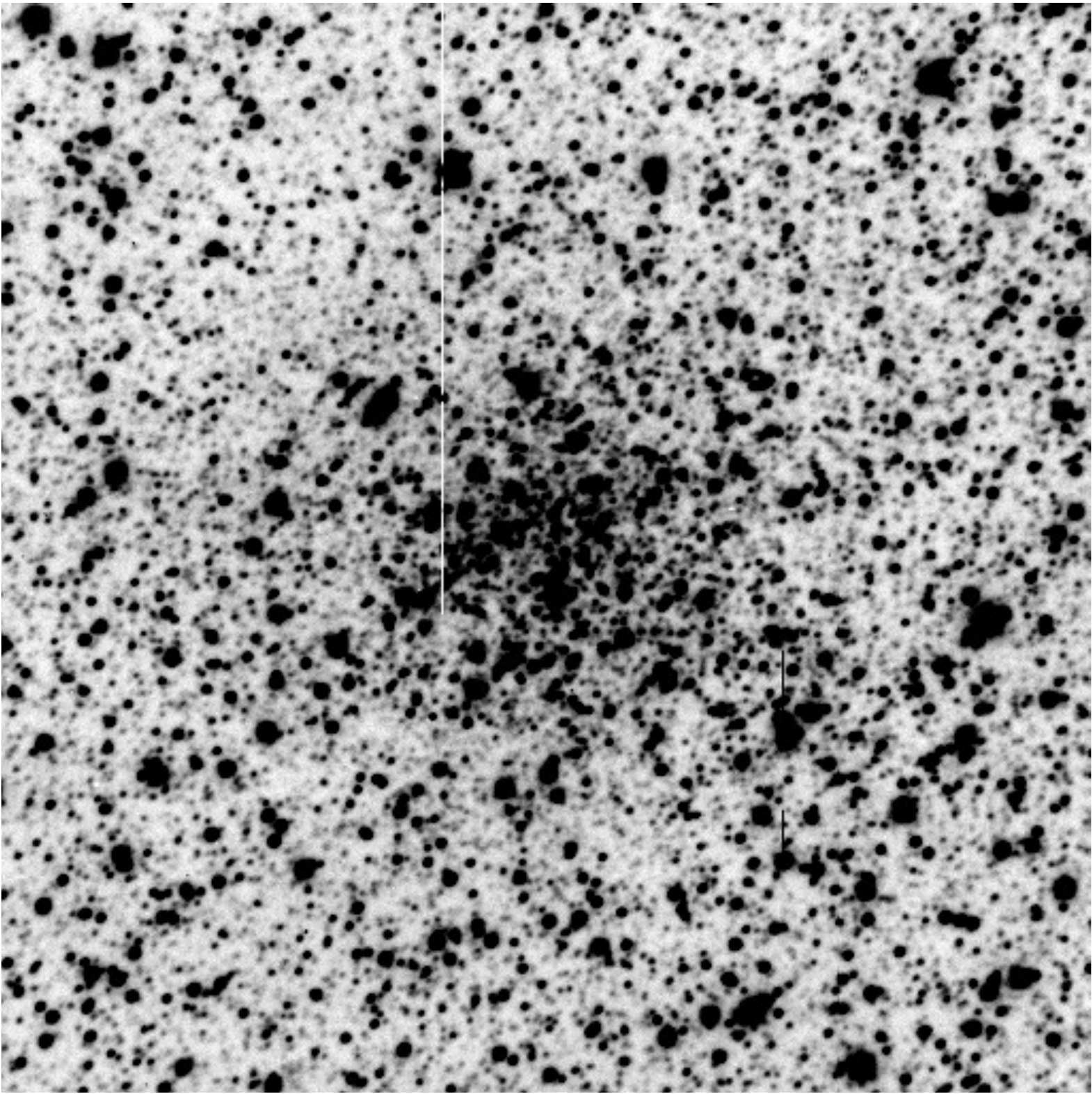}
\caption {NTT 240s I image of Djorgovski 1. North is up, east to the right.
Size is 2$\times$2 arcmin$^{2}$. }
\label{dj1}
\end{figure}

Terzan 10 and Djorg 1 I-band images, collected with EFOSC2@NTT in 2012, 
are presented in Figures \ref{ter10} and \ref{dj1}. 
Terzan 10 is very compact ($c = 0.8$) according to the  structural analysis of Bonatto \& Bica (2008).
Djorg 1 is rather loose with no evidence of post-core collapse (Trager et al. 1993).
% In a forthcoming paper we will analyse the  structures of the two clusters usi%ng the present HST material.

%-------
%--------------------------------------------------------------------

%-------
%--------------------------------------------------------------------
 
\section{Colour-magnitude diagrams and cluster parameters}

\begin{table}
\caption[1]{\label{literature}Literature values. References: 1. Ortolani et al. (1997); 2 Harris (1996, updated in 2010);
 3. Alonso-Garcia et al. (2015); 4 Geisler et al. (2018, in preparation);
  5 Ortolani et al. (1995); 6 Davidge 2000; 7 Valenti et al. (2010); 8 V\'asquez et al. (2018).
 $^a$ E(J-K)=0.86 transformed to E(B-V) using Fitzpatrick (1999) formula.  }
%\begin{flushleft}
\begin{tabular}{ll@{}r@{}r@{}r@{}r@{}r@{}}
\noalign{\smallskip}
\hline
\noalign{\smallskip}
 {\rm Cluster} & {\rm \phantom{-}E(B-V)} & \hbox{\phantom{-}\phantom{-}\phantom{-}d$_{\odot}$(kpc)}
& {\rm \phantom{-}\phantom{-}R$_{\rm GC}$(kpc)} & {\rm \phantom{-}[Fe/H]} 
& \hbox{\phantom{-}\phantom{-}\phantom{-}v$_{\rm r}$(km s$^{-1}$)} & \hbox{\phantom{-}ref.}  \\
\noalign{\smallskip}
\hline
\noalign{\smallskip}
\hbox{Terzan~10}   & 2.40$\pm$0.15 & 4.8$\pm$1    & 3.2         & $-$1.0     & ---     &  1 \\
                    & 2.40$\pm$0.15 & 5.8$\pm$1    & 2.3         & $-$1.0     &  ---    & 2 \\
                    & 1.72$^a$      & 10.3$\pm$0.2 & 2.1$\pm$0.2 & $-$1.0     & ---     & 3 \\
                    & ---           & ---          &  ----       & ---        & 208.6$\pm$3.6 & 4 \\
\hbox{Djorg~1} & 1.71$\pm$0.10 & 8.8$\pm$1    & ---         & $-$0.4     & ---    & 5 \\
                    & 1.58$\pm$0.15 & 13.7         & 5.7         & $-$1.51     & ---    & 2 \\
                    & 1.44$\pm$0.10 & ---          & ---         & $<$$-$2.0   & ---    & 6 \\
                    & 1.58$\pm$0.15 & 13.5         & 5.5         & $-$1.51     & ---    & 7 \\
                    & ---           & ---          & ---         & $-$1.36     & -358.1 & 8 \\
\noalign{\smallskip}
\hline
\end{tabular}
%\end{flushleft} 
\end{table}

Table \ref{literature} gives basic parameters taken from the literature
 for the sample clusters. 
 Only recently, metallicities, and in particular the radial velocities, of the two sample clusters were derived from  spectroscopic data.
For Djorg 1, medium-resolution spectra were obtained with FORS2@VLT in the
region of the CaII triplet lines (CaT) by V\'asquez et al. (2018). 
The CaT calibrations developed by Saviane et al. (2012) and improved
in V\'asquez et al. (2018) appear to be reliable.
For Terzan 10, Geisler et al. (in preparation) obtained high-resolution
CRIRES@VLT spectra of three stars, having derived their metallicities and radial
velocities.

\begin{figure}
\includegraphics[bb=18 144 592 718,angle=0,width=11cm]{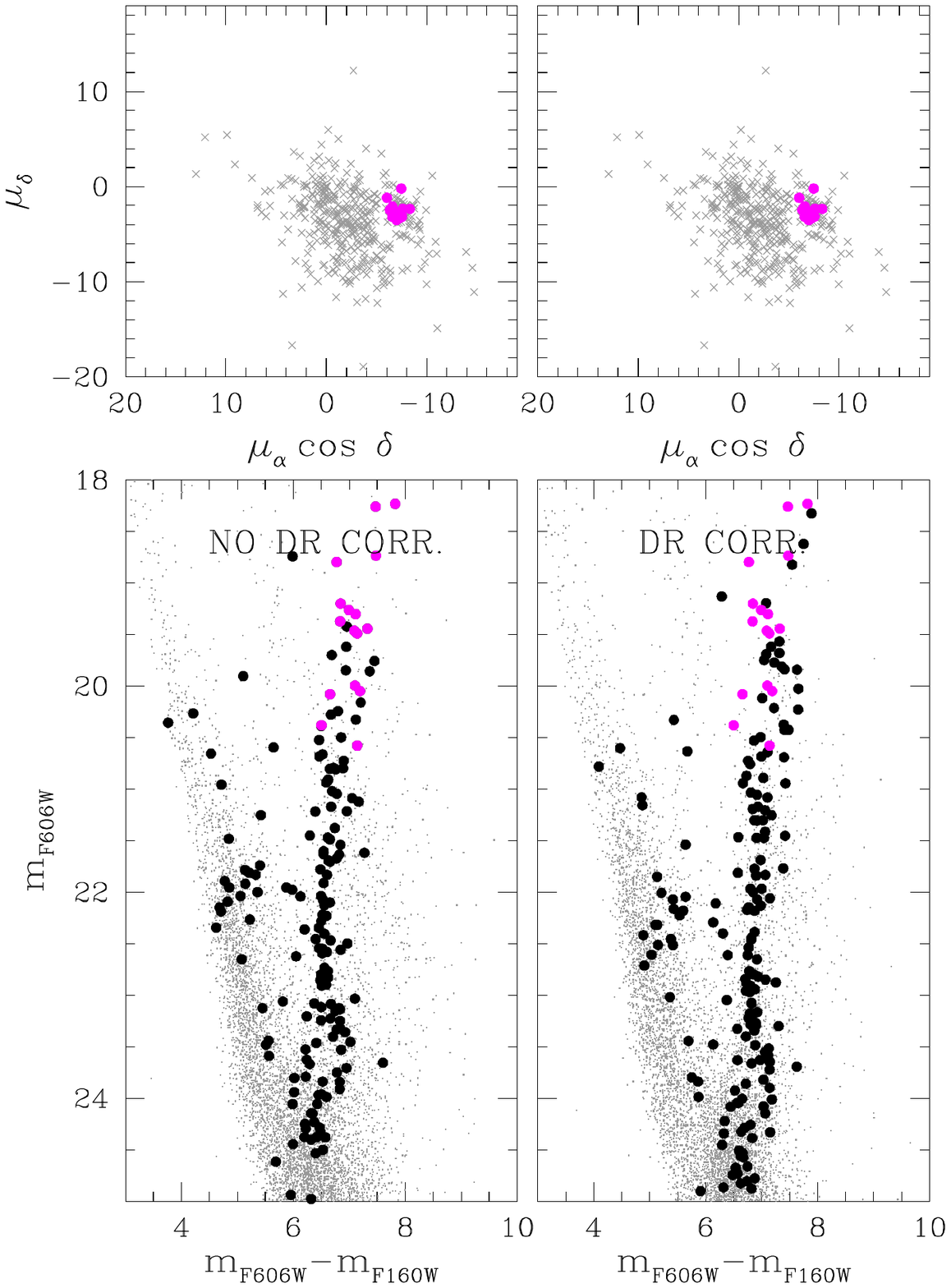}
\caption {\label{nardiello1}
 Terzan~10: upper panels show the Gaia DR2 proper motions for the stars in common between the HST catalogue and Gaia DR2; magenta dots are the stars located within $15$~arcsec from the cluster centre that also have a proper motion $<2$~mas~yr$^{-1}$ from the mean motion of the cluster (i.e. -6.96,-2.45~mas~yr$^{-1}$). Bottom panels show the  $m_{\rm F606W}$ vs. $m_{\rm F606W}-m_{\rm F160W}$ CMD before (left-hand panel) and after (right-hand panel) the differential-reddening correction. Grey points are all the stars in the HST catalogue; black points are the stars in the HST catalogue that are located within $15$~arcsec from the cluster centre; magenta points are the stars in common with Gaia DR2 catalogue that are located within $15$~arcsec from the cluster centre and have a proper motion $< 2$~mas~yr$^{-1}$ from the mean motion of the cluster.}
\end{figure}

\begin{figure}
\includegraphics[angle=0,width=11cm]{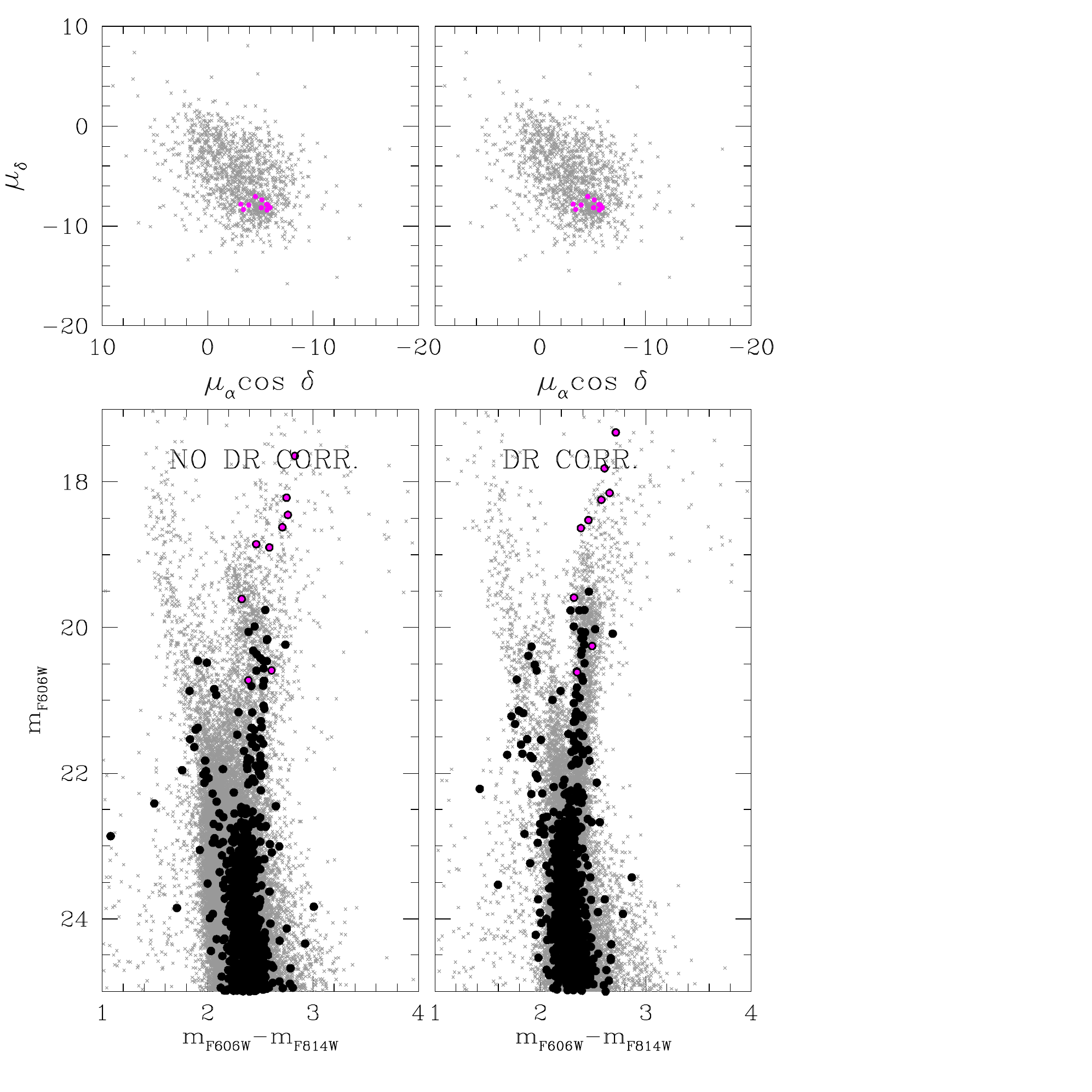}
\caption{Djorgovski~1: upper panels show the Gaia DR2 proper motions for the stars in common between the HST catalogue and Gaia DR2; magenta dots are the stars located within $8$~arcsec from the cluster centre that also have a proper motion $<2$~mas~yr$^{-1}$ from the mean motion of the cluster (i.e. -4.99,-8.32~mas~yr$^{-1}$). Bottom panels show the  $m_{\rm F606W}$ vs. $m_{\rm F606W}-m_{\rm F814W}$ CMD before (left-hand panel) and after (right-hand panel) the differential-reddening correction. Grey points are all the stars in the HST catalogue; black points are the stars in the HST catalogue that are located within $10$~arcsec from the cluster centre; magenta points are the stars in common with Gaia DR2 catalogue that  are located within $10$~arcsec from the cluster centre and have a proper motion $< 2$~mas~yr$^{-1}$ from the mean motion of the cluster.
}
\label{nardiello2}
\end{figure}

Figures \ref{nardiello1} and \ref{nardiello2} show  the Gaia DR2 proper motion
values for all stars in common between HST and Gaia (upper panels) 
in Terzan 10 and Djorg 1, respectively.
Magenta dots identify red giant branch members of both clusters, located within
 $<$10 arcsec and $<$ 20 arcsec  of the centres of Terzan 10
and Djorg 1, respectively. For Terzan 10,
 the lower panels of Fig. \ref{nardiello1}
 show m$_{F606W}$ versus m$_{F606W}$$-$m$_{F160W}$ CMDs
 for the original (left panel), and  
 differential-reddening-corrected (right panel) photometry.

 It appears that the observed CMD
(left panel) is better defined, in particular the blue horizontal
branch. This is due to the fact that differential-reddening corrections
need a well-defined cleaned sequence of bona fide member stars.
For this reason, in the subsequent analysis we use the original
photometric data for Terzan 10.

For Djorg 1, the upper panels from Fig. \ref{nardiello2} are similar to
Fig. \ref{nardiello1}; the lower panels of Fig. \ref{nardiello2} show 
 the  m$_{F606W}$ versus m$_{F606W}$$-$m$_{F814W}$ CMDs 
that correspond to the original photometry (left panel), and  
 to differential-reddening corrected CMD\ data (right panel).
 In this case the differential-reddening-corrected CMD is improved
relative to the observed one, and is therefore adopted for the analysis.

The differential-reddening corrections
were carried out following the procedures employed by
Milone et al. (2012), briefly described as follows.
 In the  $m_{\rm F606W}$ versus $m_{\rm F606W}-m_{\rm F814W}$ 
CMD of Djorg~1, and $m_{\rm F606W}$ versus $m_{\rm F606W}-m_{\rm F160W}$ CMD
 of Terzan~10, for a given target star, we selected the 40 closest cluster
 stars and measured their mean colour offset from the fiducial cluster
 sequence (along the reddening correction). This quantity is the local
 estimate of the differential reddening to be used to correct the magnitude
 of the target star.

For both clusters a blue horizontal branch is clearly detected.
This morphology together with their moderately metal-poor metallicity
indicates that the clusters should be very old, as in for example NGC 6522 (Kerber
et al. 2018a).

%-------
%--------------------------------------------------------------------

%-------
%--------------------------------------------------------------------

%-------
\begin{figure}
\includegraphics[angle=0,width=9cm]{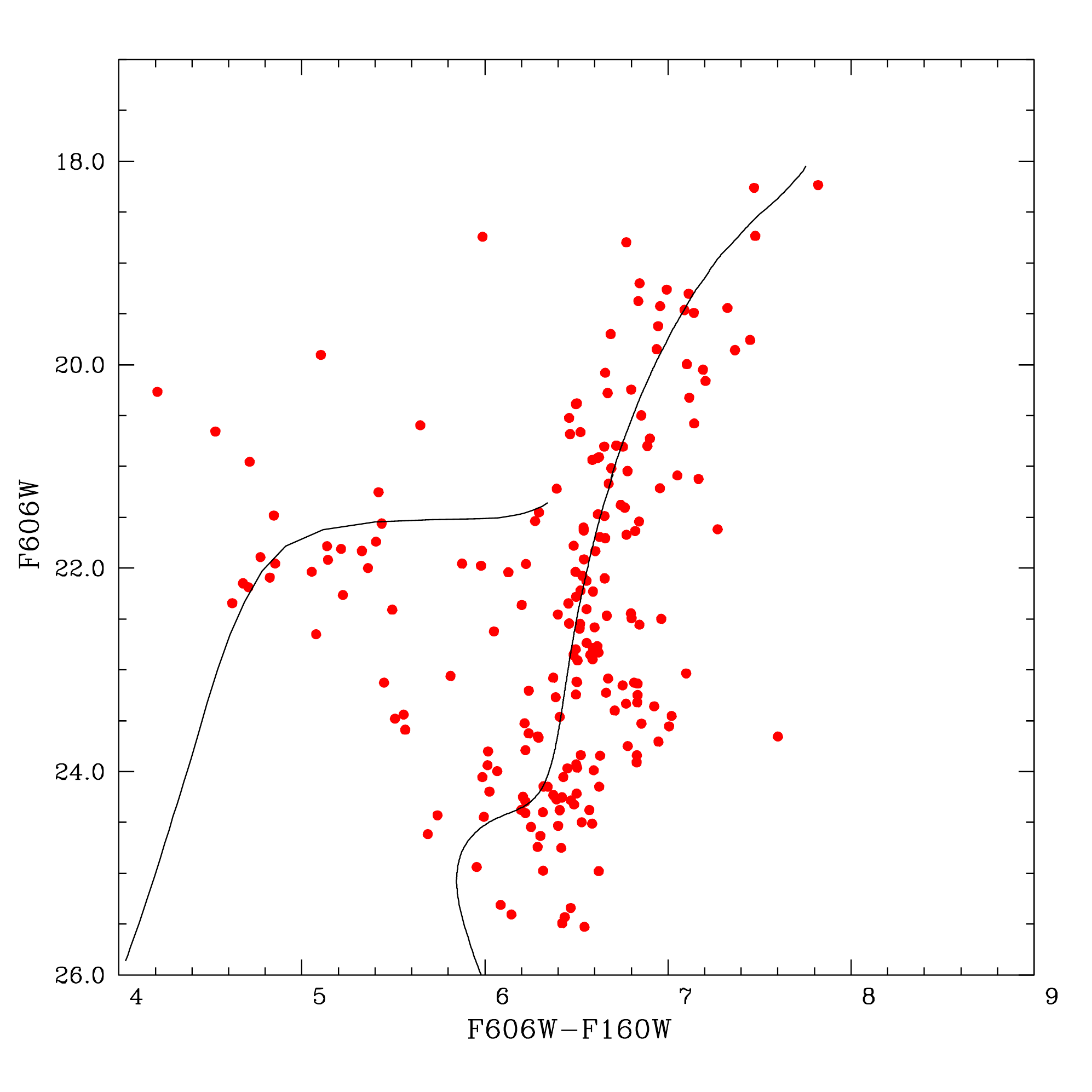}
\caption {Terzan 10:  m$_{F606W}$ vs. m$_{F606W}$-m$_{F160W}$ CMD,
 for an extraction with   1.5 arcsec $<$ r $<$ 11 arcsec.
A BaSTI alpha-enhanced isochrone of Z=0.001, 
and 13 Gyr is overplotted.
}
\label{tz10new}
\end{figure}
%--------------------------------------------------------------------

%-------
\begin{figure}
\includegraphics[angle=0,width=9cm]{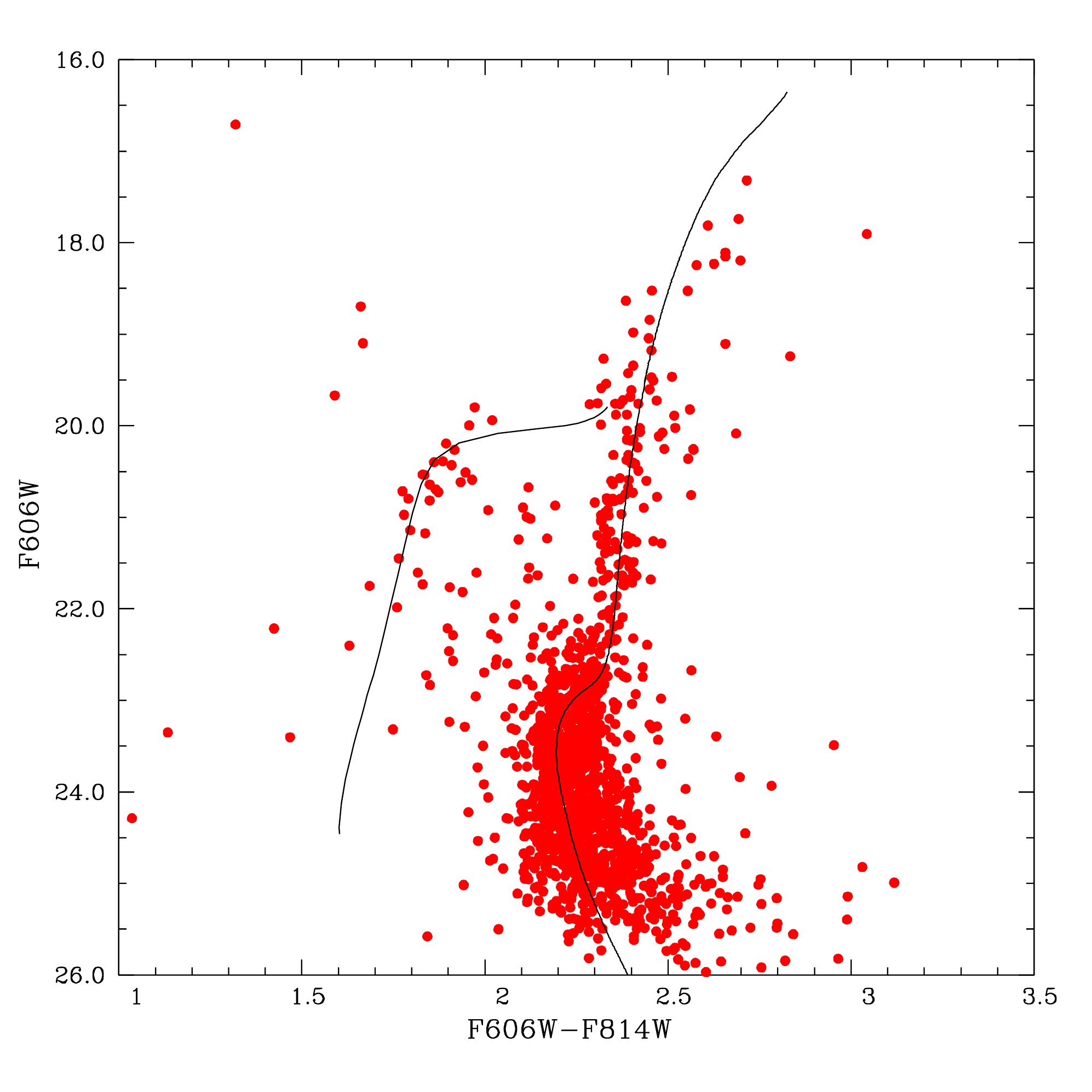}
\caption {Djorg 1:  m$_{F606W}$ vs. m$_{F606W}$-m$_{F814W}$ 
differential-reddening-corrected CMD. 
The data correspond to a central extraction
of  r $<$ 20 arcsec.
A BaSTI alpha-enhanced isochrone of Z=0.001, and 13 Gyr is overplotted.
}
\label{dj1new}
\end{figure}
%--------------------------------------------------------------------

Figures \ref{tz10new} and \ref{dj1new} show the
 m$_{F606W}$ versus m$_{F606W}$$-$m$_{F160W}$ non-differential-reddening-corrected CMDs
for Terzan 10,  and
 m$_{F606W}$ versus m$_{F606W}$$-$m$_{F814W}$ 
differential-reddening-corrected CMDs for Djorg 1.
In order to minimize the contamination by field stars and to use only well measured
 stars in the isochrone-fitting procedure, we selected stars located between
 1.5~arcsec and 11.0~arcsec (30 $<$ r $<$ 220 in pixels) from the centre of Terzan~10 (Fig. \ref{tz10new}) and within 20~arcsec (r $<$ 400 pixels) from the centre of Djorg~1 (Fig. \ref{dj1new}). These  radial interval
selections are slightly different relative to those used for Figs. \ref{nardiello1} and 
\ref{nardiello2}.

The BaSTI (Pietrinferni et al. 2004, 2006) alpha-enhanced isochrone of Z=0.001 and 13 Gyr is overplotted in both these figures.
The alpha-enhanced BaSTI isochrones were corrected for reddening-dependent 
effective temperatures, as discussed in Ortolani et al. (2017).
This correction mainly shrinks the CMDs in colours, and the fit quality
greatly improves for very reddened clusters, such as Djorg 1.

 In the derivations of cluster parameters below, 
the effect of the uncertainty on metallicity is negligible:
a change of [Fe/H]$\pm$0.1 has a minor effect on the distance. This
corresponds to a reddening difference of E(606-160)=$\pm$0.03 mag
or A(606)= $\pm$0.04, producing a distance error of $\pm$0.02 kpc. The
effect on the Horizontal Branch
(HB), corresponding to a difference in the absolute HB magnitude from the models,
 is a variation of only $\Delta$M$_{(606)HB}$= $\pm$0.02 mag.
Combining the two effects (they affect the distance modulus in the
same way) we get a total error in the distance of $\pm$0.04 kpc, which is
very small compared to other effects.
The error in the distance is then a combination of the photometric
errors, contamination, reddening and [Fe/H]; the first two are
likely dominating however.

\subsection{Terzan~10}

From the isochrone fit, we derive an apparent distance modulus of 
(m-M)$_{F606W}$ = 21.1$\pm$0.1. Measuring A$_{\rm F606W}$ = 6.02$\pm$0.05 in the
CMD, %of Fig. 1,
an absolute distance modulus (m-M)$_{0}$ = 15.08 is obtained.
A reddening E(F606W-F160W) = 4.83 converts to E(B-V)=2.17.
 We used the standard value of R$_{\rm V}$ = A$_{\rm V}$/E(B-V) = 3.1.
The corresponding visual absorption                 
A$_{\rm V}$ =   4.83/0.718 = 6.73, where the factor A$_{\rm V}$/A$_{\rm F606W}$
is given in the PARSEC isochrone site (Bressan et al. 2012).
Finally, a distance from the Sun of d$_{\odot}$ = 10.3$\pm$1 kpc is obtained. 

%E(B-V)=4.7/(3.1x0.718)=0.449x4.7=2.11
%E(B-V)=4.83/(3.1x0.718)=0.449x4.7=2.17

Alonso-Garcia et al. (2015) obtained a distance d$_{\odot}$ = 10.3 kpc by selecting
possible RR Lyrae members. Our previous analysis of Terzan 10, given in
Ortolani et al. (1997), provided E(B-V)=2.40, and a distance of Terzan 10
 to the Sun a factor two closer to us. Those data were not deep enough, 
and the horizontal branch was barely detected. The present, deeper data give 
a reliable distance based on a now clear blue HB, in very good agreement 
with Alonso-Garcia et al. (2015).

%For a metallicity Z=0.004, and age of 10 Gyr, we derive
%a distance modulus of m-M=21. Adopting E(V-I)=3.3 and
% using Dean et al. (1978): E(V-I)/E(B-V)=1.33 

\subsection{Djorgovski~1}

Table \ref{literature} shows that there are large discrepancies
in metallicity and distance measurements, but not in reddening. The aim
of the present work is to solve the distance ambiguity
in the literature, by using  deep high-resolution images and optimized CMDs from 
differential-reddening corrections.

 From the isochrone fit, we derive an apparent distance modulus of 
(m-M)$_{F606W}$ = 19.6$\pm$0.1. A measurement of A$_{\rm F606W}$ = 4.76$\pm$0.05 in the CMD
%of Fig. 1, ]
and an absolute distance modulus (m-M)$_{0}$ = 14.85 is obtained.
A reddening
E(F606W-F814W) = 1.63 converts to E(B-V)=1.66.
The corresponding visual absorption                 
is A$_{\rm V}$ = 5.16.  %where the factor A$_{\rm V}$/A$_{\rm F606W}$
%is given in the PARSEC isochrones site (Bressan et al. 2012).
Finally, we get a distance to the Sun of d$_{\odot}$ = 9.3$\pm$0.5 kpc.

\section{Orbits of Terzan 10 and Djorg 1}

For the first time, we are able to estimate the probable Galactic orbit of both clusters, Terzan 10 and Djorg 1. This is due to the combination of the proper motions from Gaia DR2 given by Vasiliev (2018), the recent radial-velocity determinations using CaT spectra that were obtained with the FORS2@VLT
(V\'asquez et al. 2018; Celeste Parisi, private communication),
 and the accurate distances calculated in this work. 

For the Galactic model, we employed an axisymmetric background that includes a S\'ersic bulge, an exponential disc generated by the superposition of three Miyamoto-Nagai potentials (Miyamoto \& Nagai 1975) following the recipe made by Smith et al. (2015), and a Navarro-Frenk-White (NFW) density profile (Navarro, Frenk \& White 1997) to model the dark-matter halo, which has a circular velocity $V_0=241$ km s$^{-1}$ at $R_0=8.2$ kpc (Bland-Hawthorn \& Gerhard 2016). For the Galactic bar, we used a triaxial Ferrer's ellipsoid, where all the mass from the bulge component is converted into a bar. For the bar potential, we consider a total bar mass of $1.2 \times 10^{10}$ M$_{\odot}$, an angle of $25^{\circ}$ with the Sun-major axis of the bar, a gradient of pattern speed of the bar of $\Omega_b= 45$, 50, and 55 km s$^{-1}$ kpc$^{-1}$, and a major axis extension of 3.5 kpc. We keep the same bar extension, even though we change the bar pattern speed. 

The integration of the orbits was made with the \texttt{NIGO} tool (Rossi 2015a), which includes the potentials mentioned above. The solution of the equations of motion is solved numerically using the Shampine--Gordon algorithm (for details, see Rossi 2015b). We adopted the right-handed, Galactocentric Cartesian system, $x$ toward the Galactic centre, and $z$ toward the Galactic North Pole. The initial conditions of Terzan 10 and Djorg 1 are obtained from the observational data, coordinates, heliocentric distance, radial velocity, and absolute proper motions given in Table \ref{tab:paraGC}. The velocity components of the Sun with respect to the local standard of rest are $(U,V,W)_{\odot}= (11.1, 12.24, 7.25)$ km s$^{-1}$ (Sch{\"o}nrich, Binney \& Dehnen 2010). In order to estimate the effect of the uncertainties associated to the clusters' parameters, we use the Monte Carlo method to generate a set of 1000 initial conditions for each cluster, taking into account the errors of distance, heliocentric radial velocity, and absolute proper motion components. With such initial conditions, we integrate the orbits forward for 10 Gyr. For each orbit, we calculate the perigalactic distance $r_{\rm min}$, apogalactic distance $r_{\rm max}$, the maximum vertical excursion from the Galactic plane $|z|_{\rm max}$, and the eccentricity defined by $e=(r_{\rm max}-r_{\rm min})/(r_{\rm max}+r_{\rm min})$.

\begin{table*}
%\centering
\caption{Parameters for the orbit integration.}
\begin{tabular}{@{}l|cc|cc@{}} 
\hline
Parameter & Value & Ref. &Value & Ref.\\
%\hline
&\multicolumn{2}{|c}{Terzan 10} & \multicolumn{2}{|c}{Djorg 1} \\
\hline
\hline
$(\alpha,\delta)_{(J2000)}$ & (18$^{\rm h}$02$^{\rm m}$57.8$^{\rm s}$, $-$26$^{\rm o}$$04'01\arcsec$ ) & This work & (17$^{\rm h}$47$^{\rm m}$28.7$^{\rm s}$, $-$33$^{\rm o}$$03'59\arcsec$)& 3 \\
$V_r$ (km.s$^{-1}$)& $208\pm3.6$ & 1 & $-358.1\pm$ 0.7& 4 \\
d$_\odot$ (kpc) &$10.4 \pm 1.0$ & This work &$9.3\pm 0.5$ & This work \\
$^{\dag}$$\mu_{\alpha} \cos \delta$ (mas yr$^{-1}$) &$-7.021\pm 0.072$ & 2 &$-5.111\pm 0.072$ & 2\\
$^{\dag}$$\mu_{\delta}$ (mas yr$^{-1})$& $-2.511\pm 0.063$ & 2 & $-8.304\pm 0.053$& 2\\

\hline
\hline
\end{tabular}
\label{tab:paraGC}
\\References --- (1) Geisler et al. (2018, in preparation); (2) Vasiliev 2018; (3) Ortolani, Bica \& Barbuy 1995; (4) V\'asquez et al. (2018). 
\\$^{\dag}$Uncertainty includes the systematic error of 0.035 mas yr$^{-1}$ (Gaia Collaboration et al. 2018b). 
\end{table*}

\begin{figure*}
\includegraphics[ scale=0.15]{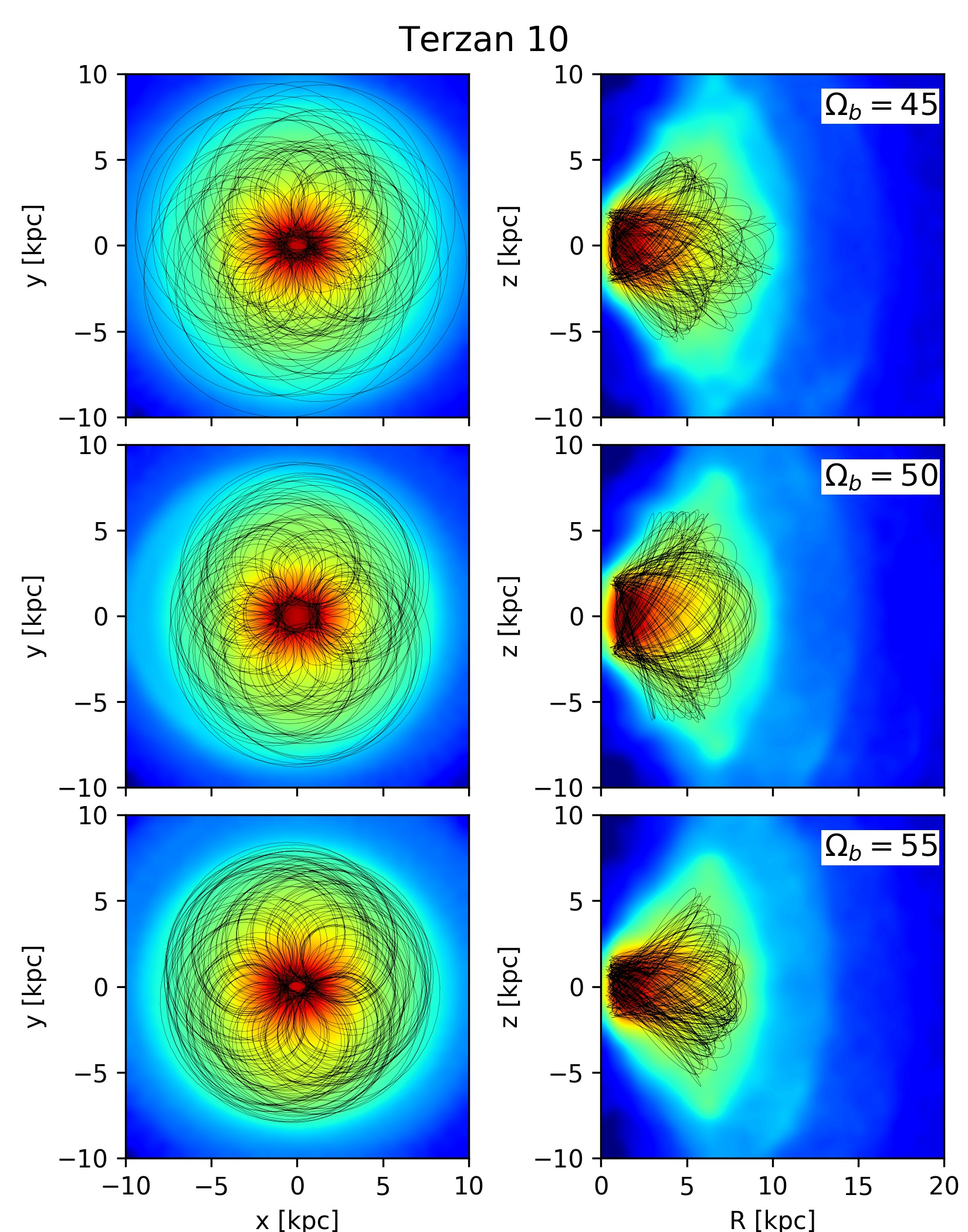}
\includegraphics[ scale=0.15]{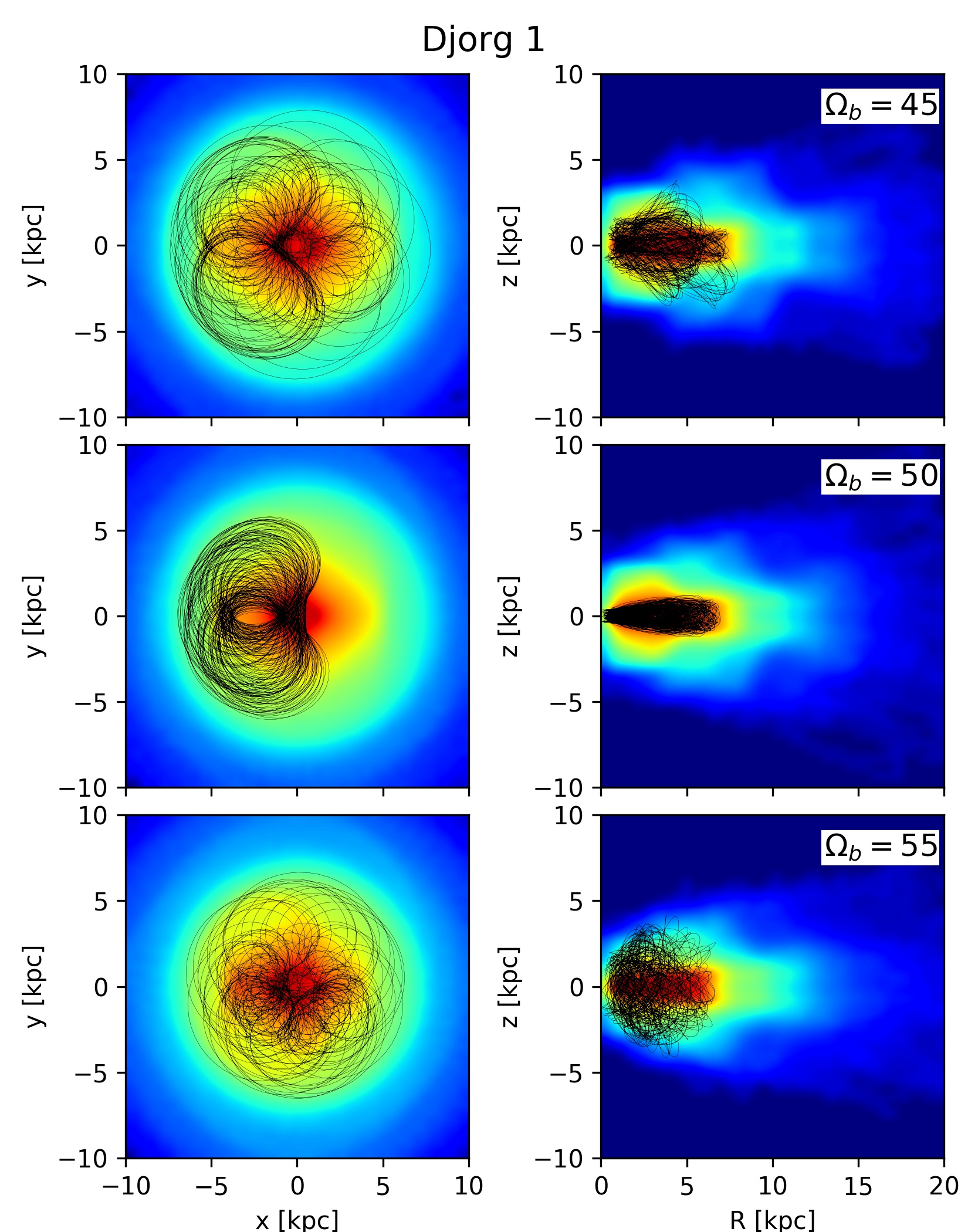}
\caption{Probability density map for the $x-y$ and $R-z$ projections of the 1000 orbits for Terzan 10 (left panels) and Djorg 1 (right panels). The orbits are co-rotating with the bar frame.
The bar pattern speed, $\Omega_b$, is given in units of km s$^{-1}$ kpc$^{-1}$.
 The red and yellow colors correspond to the larger probabilities. The black lines show the orbits using the central values presented in Table \ref{tab:paraGC}. }
\label{orbits}
\end{figure*}

The results of the orbit integration are shown in Figure \ref{orbits}, displaying the probability densities of the orbits in the $x-y$ and $R-z$ projection co-rotating with the bar. The red and yellow colours exhibit the region of the space that the orbits of Terzan 10 (left panels) and Djorg 1 (right panels) cross most frequently. The black curves are the corresponding orbit using the central values of the cluster observational parameters.

\begin{figure*}
\includegraphics[ scale=0.80]{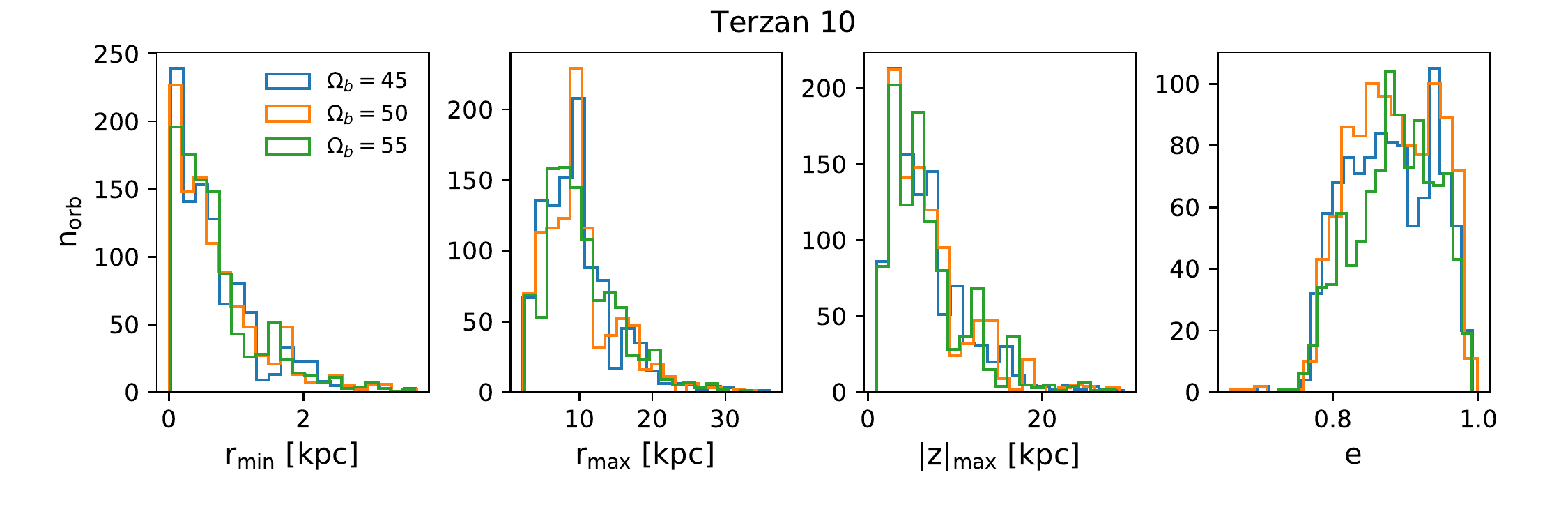}
\includegraphics[ scale=0.80]{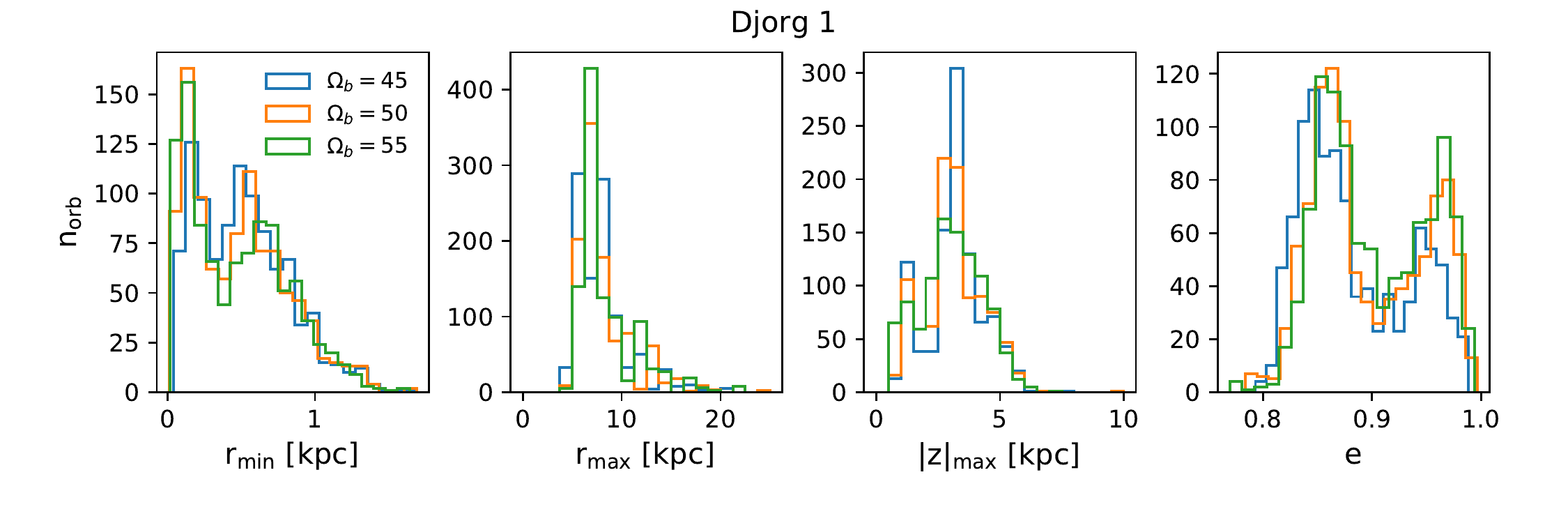}
\caption{Distribution of orbital parameters for Terzan 10 (top panels) and Djorg 1 (bottom panels), perigalactic distance $r_{\rm min}$, apogalactic distance $r_{\rm max}$, maximum vertical excursion from the Galactic plane $|z|_{\rm max}$, and eccentricity. The colours show the different angular speed of the bar, $\Omega_b$ = 45  (blue), 50 (orange), and 55 (green) km s$^{-1}$ kpc$^{-1}$.}
\label{histogram}
\end{figure*}

Distributions for the perigalactic distance, apogalactic distance, maximum vertical height and the eccentricity are presented in Figure \ref{histogram}, for Terzan 10 (top panels) and Djorg 1 (bottom panels), the different colours represent the angular velocities investigated here. The orbits of Terzan 10 have radial excursions between $\sim 0.1$ and $\sim 30$ kpc, with maximum vertical excursions from the Galactic plane between $\sim 1$ and $\sim 20$ kpc, and eccentricities $e>0.7$.  For Djorg 1, the radial excursion is between $\sim 0.1$ and $\sim 20$ kpc, with maximum vertical excursions from the Galactic plane between $\sim 0.1$ and $\sim 6$ kpc, and eccentricities $e>0.75$. The eccentricity distribution of
 Djorg 1 clearly shows a double peak that is related to the 
double peak also presented in the perigalactic distance: the
 smaller perigalactic distances with r$_{\rm min}$$\sim$2 kpc 
(left peak in first panel)
 correspond to the higher eccentricities
$e>\sim 0.9$ (right peak in fourth panel).
The variation of the angular velocity seems to have a negligible
 effect on the orbits in both clusters.

In Table 3 we present average orbital parameters of the set of orbits for Terzan 10 and Djorg 1,
where the errors provided in each column are obtained by considering the 16th and 84th percentile of the distribution. The orbital characteristics of both clusters are inconsistent with the bulge GCs that have [Fe/H]$\sim -1.0$ (P\'erez-Villegas et al. 2018), meaning that these clusters are intruders in the Galactic bulge, and
 their orbital parameters are more consistent with GCs that belong to the halo or a thick disc component.

\begin{table*}
\begin{center}
\caption{Monte Carlo average orbital parameters of Terzan 10 and Djorg 1.}
\begin{tabular}{@{}ccccc@{}} 
\hline
$\Omega_b$ & $\langle r_{\rm min}\rangle$ & $\langle r_{\rm max} \rangle$ & $\langle |z|_{\rm max}\rangle $ &$\langle e\rangle$ \\
(km s$^{-1}$ kpc$^{-1}$) & (kpc) & (kpc) & (kpc) & \\
\hline
\multicolumn{5} {c} {Terzan 10} \\
\hline
45 & $0.720^{+0.677}_{-0.390} $ & $ 9.669^{+3.893} _{-4.221} $ & $ 6.779^{+4.955}  _{-2.536}$ & $ 0.879^{+0.068}_{-0.064} $\\
50 & $0.711^{+0.745}_{-0.364} $ & $ 9.964 ^{+6.063} _{-4.029}$ & $ 6.877^{+5.490}_{-2.942} $ &$  0.884^{+0.067} _{-0.062} $\\
55 & $ 0.714^{+0.821} _{-0.357}$ & $ 10.398^{+5.928} _{-3.072}$ & $ 6.871^{+6.117} _{-2.696} $ & $ 0.886^{+0.057}_{-0.068} $ \\
\hline
\multicolumn{5} {c} {Djorg 1} \\
\hline
45 & $0.514^{+0.335}_{-0.314}$ & $7.907^ {+1.991}_{-2.034}$ & $3.178 ^{+1.113}_{-1.349}$ & $0.883 ^{+0.078}_{-0.033} $\\
50 & $0.479^{+0.349} _{-0.335} $ & $8.135^{+3.239}_{-1.513}$ & $3.144^{+1.358}_ {-1.295} $ &$ 0.896^{+0.083}_{-0.032}$\\
55 & $0.476^{+0.384} _{-0.338} $ & $ 8.348^{+4.889} _{-0.667}$ &$ 3.061 ^{+1.358}_ {-1.504}  $ & $ 0.901^{+0.074} _{-0.040} $ \\

\hline
\end{tabular}
\end{center}
\label{tab:Orbpara}
\end{table*}

\section{Discussions and Conclusions}

 Terzan 10 and Djorg 1 were selected given that they
are moderately metal-poor ([Fe/H]$\sim$$-$1.0)
 as listed in Barbuy et al. (1998, 2009),
and they are projected close to the Galactic centre. 
The aim is to identify genuine old bulge clusters, such as
NGC~6522 (Barbuy et al. 2014, Kerber et al. 2018a),
 HP~1 (Barbuy et al. 2016, Kerber et al. 2018b),
and NGC~6558 (Barbuy et al. 2018).
 Furthermore, due to difficulties
of crowding and absorption, there are very few studies of these clusters,
and discrepancies are found in literature parameters. 

Deep CMDs were obtained using HST optical/infrared filters (F606W, F160W) 
for Terzan 10,
and (F606W, F814W) for Djorg 1.
Since the field contamination is very high,
 it was only possible to correct
the CMD of Djorg 1 for differential reddening. The CMDs are deep
and accurate enough to reveal blue horizontal branches for both
clusters. 
We derived reddening and distance values that supersede
the literature uncertainties reported in Table \ref{literature}.

We estimated the absolute total magnitude for the sample clusters
by counting red giant branch stars above the horizontal branch level,
and taking into account the cluster profiles.
Based on the absolute total magnitudes of template clusters in Harris (1996, 2010 edition), for the RGB counts we used 13 GCs, especially selecting distant halo GCs so as to include as many giants as possible, according to the CMDs given by  M. Castellani \footnote{ http://gclusters.altervista.org/, by M. Castellani, INAF, Osservatorio Astronomico di Roma}. We obtain 
M$_{\rm V}^{\rm t}$ = -5.8$\pm$0.4  for both clusters,  which is fainter than
 the values of -6.35 and -6.98  for Terzan 10 and
 Djorg 1, respectively, given in Harris (1996, 2010 edition). 
These results are comparable to most Palomar clusters, which are
more luminous than ultra-faint clusters (e.g. Kim \& Jerjen 2015, Luque et al. 2016), and fainter than classical halo clusters. This faint magnitudes suggest a  mass loss along their trajectories crossing the 
bulge and disc in many orbits (Aguilar et al. 1988). 
Together with Gaia proper motions, radial velocities
from the recent literature, and our improved distances,
we were able to reach our ultimate objective of
computing their orbits.
The high values of proper motions and radial velocities
lead to apogalactic distances of about 10-20 kpc, characterising
Terzan 10 and Djorg 1 as halo clusters. Therefore,
although these clusters are projected in the central parts
of the bulge, that is, they are halo intruders.

\begin{acknowledgements}
We are grateful to Celeste Parisi and Doug Geisler for
providing the radial velocity of Terzan 10, in advance
of publication.
 SO and DN acknowledge partial support by the Universita` degli Studi
di Padova Progetto di Ateneo CPDA141214 and BIRD178590 and
by INAF under the program PRIN-INAF2014.
 APV acknowledges a FAPESP grant no. 2017/15893-1.
 BB  and  EB acknowledge  grants from  the
brazilian  agencies  CAPES - Finance code 001, CNPq and  FAPESP.

\end{acknowledgements}

%--------------------------------- References -------------

\end{document}